\input{aipcheck}

\documentclass[
    ,final            
  ]
  {aipproc}

\layoutstyle{6x9}

\begin{document}

\title{Charm Factories: Present and Future}


\author{Peter Zweber (on behalf of the BESIII Collaboration)}{
  address={University of Minnesota, Minneapolis, MN 55455 USA}
}

\begin{abstract}
The next generation tau-charm factory,
the third Beijing Electron Spectrometer (BESIII)
at the new Beijing Electron Positron Collider (BEPCII),
has begun data collection.  I discuss the flavor physics reach 
of the BESIII charm program and conclude with a discussion 
on future proposed tau-charm facilities.
\end{abstract}

\maketitle


\section{Introduction}

The current landscape for experiments running in the charm region has recently changed.  
The CLEO-c experiment ceased data collection in March 2008.  The BESIII experiment has 
begun to collect data.  I review the current and 
future experiments to be run in the charm region, in addition to selected topics in which 
the BESIII collaboration is expected to make substantial contributions to charm physics.
I focus on two aspects of the program: $D^+$ and $D_S^+$ leptonic 
decays\footnote{The use of charge conjugate modes is implied unless otherwise indicated.} 
and studies of $D^0 \rightarrow K^0_S \pi^+ \pi^-$ decays with their impact on the measurement 
of the CKM angle $\gamma/\phi_3$.  
While other measurements, e.g., semileptonic $D^0$, $D^+$, and $D_S^+$ decays, 
quantum-correlated analyses, and rare $D^0$ and $D^+$ decays, are of great interest, 
they are not discussed here.

\section{BEPCII and BESIII}

The upgraded Beijing Electron Positron Collider, BEPCII \cite{BESIIIBook}, 
is an electron-positron accelerator with separate storage rings for each beam.  
It is designed to run with 93 bunches for a maximum current of 910 mA per beam.
The center-of-mass energy ($E_{CM}$) range of the $e^+e^-$ collisions is 3.0 - 4.6 GeV.  
The design luminosity is $1\times10^{33}~{\rm cm}^{-2}~{\rm s}^{-1}$ 
for $E_{CM}$ = $M[\psi(2S)]$ and $M[\psi(3770)]$ and 
$0.6\times10^{33}~{\rm cm}^{-2}~{\rm s}^{-1}$ near $M(J/\psi)$ and $E_{CM}$ > 4 GeV.    

The BESIII detector \cite{BESIIIBook} is an approximately cylindrically symmetric detector 
that provides a solid angle coverage of $93\%$.  
The internal components consist of a 43-layer wire drift chamber (MDC), 
a time-of-flight (TOF) system with 2 layers in the barrel region and one layer for each endcap, 
and a 6272 cell CsI(Tl) crystal calorimeter (EMC).  
These components reside within a magnetic field aligned with the beam axis. 
A Muon Chamber (MUC) consists of 9 layers of resistive plate chambers 
within the return yoke of the magnet.  
The momentum resolution for charged tracks in the MDC is 0.5\% for transverse momenta of 1 GeV/$c$.  
The energy resolution for showers in the EMC is $\sim4\%$ (2.3\%) for 0.1 (1) GeV photons.  
Charged particle identification from measurements of ionization energy loss within the 
MDC and information from the TOF provides better than two standard deviations of separation 
for $\pi$ and $K$ tracks with net momenta less than 900 MeV/$c$.  
The MUC provides reconstruction efficiencies of greater than 90\% 
for track momenta in excess of 500 MeV/$c$.

The BESIII detector was commissioned last year.  
As of April 2009, 100 million $\psi(2S)$ decays have been collected.  
Table \ref{tab:projdata} lists the number of events for an average year with BEPCII 
running at design luminosities \cite{BESIIIBook}.  

\begin{table}
\begin{tabular}{ccccc}
\hline
$E_{CM}$ & BESIII             & BESII           & CLEO-c          & Type \\
(GeV)    & ($10^6$ events/yr) & ($10^6$ events) & ($10^6$ events) & \\
\hline
3.097 & 10000 & 58    & ---  & $J/\psi$         \\
3.686 & 3000  & 14    & 27   & $\psi(2S)$       \\
3.773 & 18.3  & 0.12  & 3.0  & $D^0\overline{D}^0$ \\
3.773 & 14.6  & 0.09  & 2.4  & $D^+D^-$ \\
4.010 & 0.8   & ---   & scan & $D^+_{S}D^-_{S}$   \\
4.170 & 2.7   & ---   & 0.55 & $D^+_{S}D^{\ast-}_{S}$   \\
\hline
\end{tabular}
\caption{Projected BESIII data samples per year 
as compared to previous experiments.}
\label{tab:projdata}
\end{table}

\section{Leptonic Decays of Charmed Mesons}

The helicity-suppressed leptonic decay of the $D^+$ ($D_S^+$) is proportional to the 
CKM matrix element $V_{cd}$ ($V_{cs}$) and the decay constant $f_{D^+}$ ($f_{D_S^+}$).  
The decay rate is given by  
\begin{displaymath}
B(D_{(S)}^+ \rightarrow \ell^+ \nu_{\ell}) = 
\frac{G^2_F~m_{D_{(S)}^+}\tau_{D_{(S)}^+}}{8\pi}~m^2_{\ell} 
\left(1 - \frac{m^2_{\ell}}{m^2_{D_{(S)}^+}}\right)
f^2_{D_{(S)}^+}~|V_{cd(s)}|^2,
\end{displaymath}
where $\ell = e, \mu, \tau$; $m_l$ is the mass of the lepton; 
and $m_{D_{(S)}^+}$ and $\tau_{D_{(S)}^+}$ 
are the $D^+$ ($D_S^+$) mass and lifetime, respectively.  
Using other measurements of the CKM matrix elements allow us to measure 
the corresponding decay constants.  
While it is interesting to measure the decay constants for their own merits, 
it is important to compare them to the Lattice QCD predictions 
so the predictions can be validated for use in the $B$ system.

Using an 818 pb$^{-1}$ data sample collected at $\psi(3770)$, the CLEO collaboration measured  
$B(D^+ \rightarrow \mu^+ \nu_{\mu}) = (3.82\pm0.32(stat)\pm0.09(syst))\times10^{-4}$, 
the most precise measurement to date.  
This leads to $f_{D^+} = 205.8\pm8.5\pm2.5$ MeV \cite{CLEOfD}, which is in agreement 
with the Lattice QCD prediction $f_{D^+} = 207\pm4$ MeV \cite{LQCDfDfDs}.  
With two years of running, BESIII can collect 10 fb$^{-1}$ of data at $\psi(3770)$ 
and decrease the experimental uncertainty of $f_{D^+}$ to 1.2\%.

The decay channel $D^+ \rightarrow \tau^+ \nu_{\tau}$ is yet to be observed.  
The most stringent upper limit is $B(D^+ \rightarrow \tau^+ \nu_{\tau})$ < 0.12\% at 90\% 
confidence level \cite{CLEOfD} using only $\tau^+ \rightarrow \pi^+ \overline{\nu}_{\tau}$.  
The Standard Model (SM) predicts $B(D^+ \rightarrow \tau^+ \nu_{\tau})$ = 0.1\%.  
With 10 fb$^{-1}$ of $\psi(3770)$ data, BESIII expects to observe about 1400 
$D^+ \rightarrow \tau^+ \nu_{\tau}$ events reconstructed in the decay modes 
$\tau^+ \rightarrow \pi^+ \overline{\nu}_{\tau}$, $\pi^+ \pi^0 \overline{\nu}_{\tau}$, 
and $\pi^+ \pi^- \pi^+ \overline{\nu}_{\tau}$.  
An observation of $D^+ \rightarrow \tau^+ \nu_{\tau}$ 
can test the SM prediction for the decay rate, 
search for CP violation in decays of $D^-$ and $D^+$, 
and provide an additional measurement of $f_{D^+}$.

The current situation for $f_{D_S^+}$ is more interesting.  Recent measurements 
of the $D_S^+ \rightarrow \mu^+ \nu_{\mu}$ and $D_S^+ \rightarrow \tau^+ \nu_{\tau}$ 
decay rates find $f_{D_S^+} = 261.2\pm6.9$ MeV \cite{CLEOfDs}.  
This is 2.6 standard deviations larger than the Lattice QCD prediction 
of $f_{D_S^+} = 241\pm3$ MeV \cite{LQCDfDfDs}.  
With 6 fb$^{-1}$ of data collected at $E_{CM} = 4170$ MeV, 
corresponding to two years of data at the peak cross section of 
$e^+e^- \rightarrow D_S^{\ast+} D_S^{-}$, 
BESIII expects to measure $B(D_S^+ \rightarrow \mu^+ \nu_{\mu})$ and 
$B(D_S^+ \rightarrow \tau^+ \nu_{\tau})$ with experimental uncertainties of 2.7\% and 2.6\%, respectively.  
This would lead to an uncertainty on $f_{D_S^+}$ of 0.9\% and, 
using the $f_{D^+}$ measurement above, on $f_{D_S^+}/f_{D^+}$ of 1.5\%.  
BESIII also expects to measure the ratio 
$B(D_S^+ \rightarrow \tau^+ \nu_{\tau})/B(D_S^+ \rightarrow \mu^+ \nu_{\mu})$ 
to a precision of 1.5\%, improving the experimental precision by a factor of 4.

BESIII may be sensitive to observing radiative leptonic decays. 
Predictions of $B(D^+ \rightarrow \gamma~\ell^+ \nu_{\ell})$ and 
$B(D_S^+ \rightarrow \gamma~\ell^+ \nu_{\ell})$, where $\ell = e,\mu$,  
are in the range $(0.1-8.2)\times10^{-5}$ \cite{DlngPred} and 
$(0.1-9.0)\times10^{-4}$ \cite{DlngPred,DDslngPred}, respectively.  
Using the data samples collected at $E_{CM} = M[\psi(3770)]$ and 4170 MeV described above, 
BESIII will be sensitive to the $D^+$ ($D_S^+$) decay rates at the level 
of $1.1\times10^{-5}$ ($1.4\times10^{-4}$). 

\section{$D^0 \rightarrow K^0_S \pi^+ \pi^-$ and Impact on $\gamma/\phi_3$}

The most precise measurements of the angle $\gamma$ are from 
$B^{\pm} \rightarrow D(K^0_S \pi^+ \pi^-) K^{\pm}$ decays, where $D = D^0$ or $\overline{D}^0$.  
Using 383 million $B\overline{B}$ decays the BaBar collaboration measured 
$\gamma = [76^{+23}_{-24}(stat)\pm5(syst)\pm5(model)]^{\circ}$ \cite{BaBarGamma}, 
while the Belle collaboration determined a preliminary value of 
$\gamma = [76^{+12}_{-13}(stat)\pm4(syst)\pm9(model)]^{\circ}$ \cite{BelleGamma} 
from 657 million $B\overline{B}$ decays.  
The model uncertainties arise from the isobar model analysis 
of flavor-tagged $D \rightarrow K^0_S \pi^+ \pi^-$ decays from continuum-produced 
$D^{\ast\pm} \rightarrow D \pi^{\pm}$ events.

Various authors \cite{Giri2003,Bondar} have proposed to remove 
the model dependence by performing binned analyses of 
the $D \rightarrow K^0_S \pi^+ \pi^-$ Dalitz plot.  
Using an 818 pb$^{-1}$ $\psi(3770)$ data sample, CLEO \cite{CLEOGamma} 
has performed a binned Dalitz plot analysis using the method 
suggested by Bondar and Poluektov \cite{Bondar}.  
They measured the strong relative phase in eight bins.  
They also determined that the model uncertainty in $\gamma$ will be about $1.7^{\circ}$ 
based on a toy MC study.  
With a 10 fb$^{-1}$ $\psi(3770)$ data sample, BESIII expects to decrease the 
model dependence to less than $1^{\circ}$ using a procedure similar to Ref. \cite{CLEOGamma}.

\section{Future Facilities}

While BESIII is the only running experiment with high sensitivity to charm physics, 
numerous other experiments expect to contribute to this area in the coming years.  
The LHCb experiment  expects to collect $10^8$ events per 2 fb$^{-1}$ in the decays 
$D^{\ast\pm} \rightarrow D(\pi^+\pi^-, K^+K^-, K^{\pm}\pi^{\mp}) \pi^{\pm}$ \cite{LHCb}. 
The proposed Super B experiments have projected instantaneous luminosities on the 
order of $10^{35}$ cm$^{-2}$ s$^{-1}$ at $E_{CM}$ =  4 GeV \cite{SuperBReports}.  
Future tau-charm factories are also being considered in Russia and Turkey.

A program for a tau-charm factory at the Budker Institute of Nuclear Physics 
in Novosibirsk, Russia, is being developed \cite{Novosibirsk}.  The design luminosity is 
greater than $10^{35}$ cm$^{-2}$ s$^{-1}$ at $E_{CM}$ = 4 GeV 
by using Crab Waist collision technology \cite{CrabWaist} in a two storage ring configuration.  
The injection tunnels already exist, and the tunnels for the 2 GeV linear accelerator (linac) 
and experimental halls on the storage rings are ready.  
A technical design report is expected by the end of 2010 with initial operations planned for 2015. 

The Turkish government is considering building a general experimental facility near Ankara, 
with the official site to be determined by 2012.  
The proposed Turkic Accelerator Complex \cite{TAC} will contain a synchrotron light source, 
free electron laser, a GeV proton accelerator, and a storage ring facility with linac injection 
for $e^+e^-$ collisions in the charm energy region.  
It is also being designed to achieve luminosities of $10^{35}$ cm$^{-2}$ s$^{-1}$ 
at $E_{CM}$ = 4 GeV by using Crab Waist collision technology.  
Proposals for asymmetric beam energies are being investigated.  
Interaction region design studies are in progress, 
a technical design report is expected by 2011, 
and the facility is expected to be built before 2020.

\section{Conclusion}

The BESIII experiment is starting to collect large samples of $J/\psi$ and $\psi(2S)$ data.  
In the near future, samples will be collected in order to begin its charm physics program.



\end{document}